\documentclass[aps,prl,reprint,floatfix,amsmath,amssymb, footinbib, superscriptaddress]{revtex4-2}

\usepackage{lipsum}
\usepackage{graphicx}
\usepackage{epstopdf}
\usepackage[dvipsnames]{xcolor}
\usepackage{hyperref}
\usepackage{bm}
\usepackage{printlen}
\usepackage{units}
\usepackage{braket}
\usepackage{xcolor}

\hypersetup{
        colorlinks,
   citecolor=blue,
   filecolor=blue,
   linkcolor=blue,
   urlcolor=blue,
   breaklinks=true}

\renewcommand{\vec}[1]{\mathbf{#1}}

\newcommand{\Trace}{\text{Tr}\,}

\renewcommand{\Re}{\text{Re}}
\renewcommand{\Im}{\text{Im}}
\newcommand{\eps}{{ \varepsilon}}

\newcommand{\GG}{{G}}
\newcommand{\MM}{\mathcal{V}}

\hypersetup{hidelinks}

\begin{document}

\title{Noise probing of topological band gaps in dispersionless quantum states
}

\author{Alexander Kruchkov}

\affiliation{Institute of Physics, {\'E}cole Polytechnique F{\'e}d{\'e}rale de Lausanne,  Lausanne, CH 1015, Switzerland, and Branco Weiss Society in Science, ETH Zurich, Zurich, CH 8092, Switzerland}

\affiliation{Department of Physics, Harvard University, Cambridge, Massachusetts 02138, USA}

\author{Shinsei Ryu}

\affiliation{Department of Physics, Princeton University, Princeton, New Jersey 08544, USA}

 \date{\today}

\begin{abstract}
We uncover a useful connection between the integrated current noise $S(\omega)$ and the topological band gap in dispersionless quantum states, $\int d \omega [ \mathcal S^{\text{flat}}_{xx} + \mathcal S^{\text{flat}}_{yy} ] = C e^2 \Delta^2$  (in units $\hbar$$=$$1$), where $C$ is the Chern number, $e$ is electric charge, and $\Delta$ is the topological band gap.
This relationship may serve as a working principle for a new experimental probe of  topological band gaps in flat band materials. Possible applications include moir{\'e} systems,  such as twisted bilayer graphene and twisted transition metal dichalcogenides,  where a band gap measurement in meV regime presents an experimental challenge. 
\end{abstract}

\maketitle

The noise is the signal: Shot noise, arising from the graininess of electric charge, 
has proven to be a useful experimental observable, enabling direct probing of underlying properties of quantum systems \cite{Landauer1993}. As an example, quantum shot noise has been used as a \textit{direct probe} of the fractional electric charge in the fractional quantum Hall effect (FQHE) \cite{Kane1994,FQHcharge1,FQHcharge2,FQHEcharge3}, an observation that validated the existence of anyonic statistics relevant to quantum computing \cite{Nayak2008}. 
While noise in quantum systems can have different origins,  entanglement is one of them \cite{Klich2009}. Upon measurement with time-resolved external tools the quantum state of entangled entities responds in uncertain (noisy) manner. Such noise originates not from the external source, but rather from the internal source, entanglement between the particles constituting the interacting system. This mechanism, in particular relevant to the topological systems, remains relatively little explored.

As it became clear from the early days of solid state physics, the notion of the band gap in the electronic spectrum is crucial for understanding the material properties. At the same time, precise experimental determination of the gap magnitude still remains a challenge, even more so when one needs to measure at the millielectron-Volt (meV) scale. For example, classical X-ray photoelectron spectroscopy (XPS) has typical resolution of hundreds of meV \cite{Greczynski2023}, with similar values for the inverse photoelectron spectroscopy (IPES) \cite{Budke2007}, putting these methods clearly aside from the gap studies in materials with flat bands, such as twisted graphene multilayers and transition metal dichalcogenides \cite{Cao2018, Park2022, Zhang2022, Tian2023, Cai2023, Park2023, Zeng2023,Xu2023}. More advanced techniques, like angle-resolved photoemission spectroscopy (ARPES) and scanning tunnelling spectroscopy (STS) can potentially offer energy resolution of the order of meV and even better, but are rather sensitive to the sample quality and require additional care in terms of the surface preparation \cite{Tsuda2005,Hanaguri2012, Kushnirenko2020}. Additionally, data interpretation for every given experimental technique can hold some intrinsic ambiguities, often leading to the apparent disagreement of the results across different measurements of the same system. Such challenges instigate novel proposals for the band gap measurement in quantum materials with flat bands and narrow band gaps.

In this paper, we propose a method for probing topological band gaps in quantum materials with narrow bands by using frequency-resolved current noise measurements at low temperatures, a standard technique in modern solid-state labs \cite{Clerk2010}. Our approach takes advantage of the nontrivial quantum geometry (Fubini-Study metric) of electrons, indicating that the probe of the topological band gaps is a consequence of the entanglement of electronic orbitals. While we highlight the case with two flat Chern bands for clarity, our derivations, implemented using Kubo linear response theory at finite temperature, are applicable to an arbitrary number of electronic orbitals, and extend to Bloch topology of various nature, e.g. Euler Bloch bands. Therefore, our proposal could be useful for more accurate measurements of narrow band gaps, which are characteristic of flat-band materials such as magic-angle twisted bilayer graphene.

\textit{Approach and methodology}.---
Within the scope of this paper, we use the field-theoretical electric transport formalism of Kubo \cite{Kubo1957}, implemented with Matsubara technique \cite{Matsubara1955} to address the \textit{finite temperature measurements}. Our focus is the current noise $ \mathcal S$, defined as the current-current correlator in the \textit{real frequencies} $\omega$, 
in the symmetrized form \cite{Lesovik1989, Kane1994}
\begin{align}
  \mathcal S_{ij} (\omega)  =   \int  dt \, e^{ i \omega t}  \langle \,    \left\{ J_i (t),   J_j (0) \right\}  \rangle, 
  \label{current-noise}
\end{align}
where $\vec J$ is the current operator,  and $i,j = x,y$ are spatial coordinates; below we work with two-dimensional (2D) systems (we use $\hbar$$=$$k_B$$=$$1$ for subsequent calculations). 
We define the current operator in the conventional way,
\begin{align}
\vec J = e  \sum_{\vec k}  c^{\dag}_{\vec k}  \frac{\partial \mathcal H} {\partial \vec k}  c^{}_{\vec k} .
\label{current}
\end{align}
In this equation, $ \mathcal H$ represents the Hamiltonian of the system, and $c^{\dag}_{\vec k}$ and $c^{}_{\vec k}$ are the creation and annihilation operators of quasiparticles with momentum $\vec k$, and the summation extends over the entire Brillouin zone (BZ).
Starting from definitions, \eqref{current-noise}-\eqref{current}, we use the Kubo formalism in Matsubara representation. Upon implementation of Wick's theorem, the quantum noise in imaginary time $\tau$  reads
\begin{align}
  \mathcal S^{\pm}_{ij} ( \tau)  =  i e^2 \sum_{\vec k} \Trace \GG_{\vec k}(\pm \tau )  \MM_i \,  \GG_{\vec k}( \tau) \MM_j, 
\end{align}
where $\MM_i $$\equiv$${\partial \mathcal H} /{\partial  k_i}$, $\GG_{\vec k} (\tau)$ is the quasiparticle propagator,  and prefactor $i$ comes from Wick's rotation to imaginary time.   To evaluate $S(\omega)$ we need to consider both contributions $ \mathcal S^{+}_{ij} ( \tau)$ and $ \mathcal S^{-}_{ij} ( \tau)$. We first focus on $ \mathcal S^{-}_{ij} ( \tau)$.
Proceeding to Matsubara transform,  
 $
 \GG (\tau) = \frac{1}{\beta} \sum_{i \omega'_n} e^{-i \omega'_n \tau}  \GG(i \omega'_n) 
$,  we obtain
\begin{align}
 \mathcal S^{-}_{ij} (i \omega_0) =   \frac{i e^2}{\beta} \sum_{\vec k}  \sum_{i \omega'_n}
\Trace \GG_{\vec k} ( i \omega'_n - i \omega_0) \vec \MM_i  \GG_{\vec k} ( i \omega'_n ) \MM_j,  
\end{align}
where $\beta$ is inverse temperature ($\beta$$=$$ 1/T$),   and $\omega_n$ are Matsubara frequencies \cite{Mahan}. Here $i \omega'_n$ are fermionic Matsubara freqencies, and $i \omega_{0}$ are bosonic. 
In what follows below,  we operate with dimensionless quantities defined as $\mathcal S(\omega) =  i e^2 \tilde{\mathcal S}(\omega)$, and restore dimensional units at the end of calculation.   Proceeding to the analytical continuation of the quasiparticle propagators, we find
\begin{align}
\tilde{  \mathcal S}^{-} _{ij} (i \omega_0) = \frac{1}{ \beta} \sum_{\vec k, i \omega'_n} \iint \limits_{-\infty}^{+\infty} d \omega_1 d \omega_2
\frac{ \Trace [ A _{\omega_1}  \MM_i \, A_{\omega_2}  \MM_j]}{(i \omega_n - i \omega_0 - \omega_1) (i \omega_n - \omega_2)  } ,
\label{noise-1}
 \end{align}
where $\omega_{1,2}$ are \textit{real frequencies} and $A(\omega)$$\equiv$$A_{\omega}$ is the spectral function,  related to the  Green's functions $A(\omega)$$=$$- \frac{1}{\pi} \text{Im} G^R (\omega)$. All quantities under the trace operator assume the $\vec k$-dependence; the trace is taken in the band basis.   

\textit{Frequency-resolved current noise in topological materials}.---To apply expression \eqref{noise-1} to real frequencies, we take advantage of the residue theorem, replacing the summation over discrete Matsubara frequencies with a contour integral. This method brings the benefit of isolating the contributions from the residues within the contour, which  leads to
\begin{align}
 \frac{1}{\beta} \sum_{i \omega'_n} 
\frac{1}{(i \omega_n - i \omega_0 - \omega_1) (i \omega_n - \omega_2)  }  = \frac{f(\omega_1) - f(\omega_2)}{\omega_1 - \omega_2 + i \omega_0} ,
 \end{align}
where $f(\omega)$ is Fermi-Dirac distribution function,  and we have taken into account that $\omega_0$ is bosonic frequency. As a result, the equation for the current noise at the imaginary axis $i \omega_0$ simplifies to
\begin{align}
\tilde { \mathcal S}^{-}_{ij}  (i \omega_0)  = \sum_{\vec k} \iint \limits_{-\infty}^{+\infty} d \omega_1 d \omega_2 [f(\omega_1) - f(\omega_2)] 
\frac{ \Trace[A_{\omega_1} \MM_i \, A_{\omega_2} \MM_j ]}{\omega_1 - \omega_2 + i \omega_0} .
 \end{align}
 We proceed by computing the two integrals individually, designating them as $\mathcal S^{-} = \mathcal S^{(1)} + \mathcal S^{(2)}$. Upon performing an analytic continuation to the real axis $i \omega_0 \to \omega + i \delta$, the first contribution to the current noise becomes
\begin{align}
\tilde {\mathcal S}^{(1)}_{ij}  (\omega)  =  \sum_{\vec k} \iint \limits_{-\infty}^{+\infty} d \omega_1 d \omega_2 f(\omega_1)  \frac{ \Trace  [A_{\omega_1} \MM_i \,  A_{\omega_2} \MM_j]}{\omega_1 - \omega_2 +  \omega + i \delta} ,
 \end{align}
This expression can be further simplified upon integrating over $\omega_2$,  and using  Kramers–Kronig relationships, 
\begin{align}
\tilde{\mathcal S}^{(1)}  _{ij}(\omega)  =    \sum_{\vec k} \int \limits_{-\infty}^{+\infty} 
d \omega_1  f(\omega_1)  \, \Trace  [ A_{\omega_1} \MM_i \,  \GG^R_{\omega_1 +  \omega}  \MM_j  ]. 
 \end{align}
The second contribution to the current noise $\tilde{\mathcal S} (\omega)$ can be expressed as 
\begin{align}
\tilde{\mathcal S}^{(2)} _{ij} (\omega)= -   \sum_{\vec k} \iint \limits_{-\infty}^{+\infty} d \omega_1 d \omega_2 f(\omega_2)  \frac{   \Trace [A_{\omega_1} \MM_i \, A_{\omega_2} \MM_j]}{\omega_1 - \omega_2 +  \omega + i \delta} .
 \end{align}
This expression can be further refined by integrating over $\omega_1$ and invoking  Kramers–Kronig  relationships,
\begin{align}
\tilde{ \mathcal S}^{(2)}_{ij}  (\omega) =   \sum_{\vec k} \int \limits_{-\infty}^{+\infty} 
 d \omega_2 f(\omega_2)  \,  \Trace [
 \GG^A_{\omega_2 - \omega}  \MM_i \,  A_{\omega_2} \MM_j]. 
 \end{align}
 The similar expression follow for $\tilde{ \mathcal S}^{+}_{ij}  (\omega)$. 
 Hence,  in the \textit{real frequency} representation, the total expression for the current noise at \textit{finite temperatures} is  given by
  \begin{align}
\mathcal S_{ij} (\omega) =    i e^2 \sum_{\vec k} \int \limits_{-\infty}^{+\infty} 
 d \omega' f(\omega')  \, \mathcal L_{ij} (\omega, \omega').
 \label{current-noise-gen}
 \end{align}
 The noise kernel $\mathcal L_{ij} (\omega, \omega')$, defined through propagators, is given in the symmetric representation
\begin{align}
\mathcal L_{ij} (\omega, \omega')  \equiv   \Trace  [ & A_{\omega'} \MM_i \GG^{\rm R}_{\omega' +  \omega}  \MM_j 
 + \GG^{\rm A}_{\omega' - \omega}  \MM_i \, A_{\omega'} \MM_j 
 \nonumber
 \\
 + &  A_{\omega'} \MM_i \GG^{\rm R}_{\omega' -  \omega}  \MM_j 
 + \GG^{\rm A}_{\omega' + \omega}  \MM_i \, A_{\omega'} \MM_j ] .
 \label{noise-core}
\end{align}
Here $ \GG^{\rm R,A} ( {\omega} ) $$\equiv$$ \GG^{\rm R,A} _{\omega} $ represent the retarded (R) and advanced (A) propagators at frequency $\omega$.

Formula \eqref{current-noise-gen} for the frequency-resolved current noise,  together with the definition of the noise kernel \eqref{noise-core},  is the key result of our paper. 
This formula \eqref{current-noise-gen} is applicable to topological systems assuming the trace can be taken in the band basis, and operators $\mathcal V_i$ are written in Haldane's prescription \cite{Haldane2004,Kruchkov2023}, 
\begin{align}
\boldsymbol{\mathcal V}_{nm} =     \vec v_{n \vec k} \delta_{nm}  + \Delta_{nm} (\vec k)  \, \langle u_{n \vec k} | \partial_{\vec k}  
u_{m\vec k} \rangle  . 
\label{velocity}
\end{align}
In this context, $\vec v_{n \vec k} = {\partial_{\vec k} \varepsilon_{n \vec k}}$ designates the conventional quasiparticle velocity within the electronic band $u_{n \vec k}$. $\Delta_{nm} (\vec k) = \varepsilon_{n \vec k} - \varepsilon_{m \vec k}$ provides the momentum-dependent gap function, reflecting the energy difference between the $n$th and $m$th bands of a multiorbital system. The second element on the right-hand side (RHS) in formula \eqref{velocity} corresponds to the Berry-induced velocity \cite{Haldane2004}; see also Ref. \cite{Blount1962}. Notably, in an ideal flat band, the Fermi velocity nullifies at all points of the Brillouin zone (BZ),  $\vec v_{n \vec k} \equiv 0$.
Therefore, for the ideal flat bands, our attention centers on the \textit{interband} contributions,
\begin{align}
\boldsymbol{ \MM}^{\rm  flat}_{ nm} \equiv \Delta_{nm} (\vec k)  \, \langle u_{n \vec k} | \partial_{\vec k}  
u_{m\vec k} \rangle  , 
\end{align}
which can be used for evaluating current noise in the limit of perfectly flat bands. 
Thereafter, our result \eqref{current-noise-gen} extends the research of Neupert-Chamon-Mudry \cite{Neupert2013} to the case of finite temperatures,  arbitrary band dispersion (including perfectly flat bands), and moderate interactions which sustain the \textit{quasiparticle poles}.  
Formula \eqref{current-noise-gen} serves as the cornerstone of our subsequent analysis pertaining to topological flat bands and associated band gaps. 

\textit{Topological flat bands.}---Our subsequent discussion concentrates on flat topological bands in a multiorbital electronic system. We retain the $\eps (\vec k)$ dependence to facilitate generalization to dispersive bands and account for approximations related to real-world flat bands, if needed. 
The impact of moderate electronic interactions preserving the quasiparticle poles will be examined. Although the formula \eqref{current-noise-gen} accommodates finite-temperature effects, the thermal noise is not a subject of our discussion, and we now restrict our consideration to the low-temperature limit $T \to 0$,
 \begin{align}
\mathcal S_{ij} (\omega) =    i e^2  \sum_{\vec k} \int \limits_{-\infty}^{\varepsilon_F} 
 d \omega'    \, \mathcal L_{ij} (\omega, \omega'),
 \label{noise_T=0}
 \end{align}
where $\varepsilon_F$ represents the Fermi energy positioned within the band gap. We assume the electronic bands to be well-resolved in the presence of moderate interactions (quasiparticle poles are well defined),  and thereafter the trace in noise kernel $\mathcal L_{ij}$ can be evaluated in the band basis.
Following formula \eqref{noise_T=0}, the frequency-dependent current noise for a system with at least one  flat band at low temperature $T\to 0$ is given by 
\begin{align}
\mathcal S^{\rm flat}_{ij} (\omega)  = e^2 \sum_{\vec k} \sum_{n} \sum_{m \ne n} \Delta^2_{nm} (\vec k)   \lambda^{nm}_{\vec k} (\omega, \varepsilon_F)   \mathfrak G^{nm}_{ij} (\vec k) ,
\label{noise-topo}
\end{align}
where $\vec k$-dependence in $\Delta^2_{nm} (\vec k)$ takes into account that other $m$$\ne $$n$ bands may or may not be flat,  and noise function is defined as $ \lambda^{nm}_{\vec k} (\omega, \varepsilon_F)   = i \int \limits_{-\infty}^{\varepsilon_F}  d \omega'  \alpha^{nm}_{\vec k} (\omega, \omega') $, with 
\begin{align}
& \alpha^{nm}_{\vec k} (\omega, \omega')   =   A_n (\omega',  \vec k) \left[  \GG^R_m (\omega' +  \omega,  \vec k) +  \GG^R_m (\omega' -  \omega,  \vec k)   \right ]
\nonumber
\\
 &  +  \left [ \GG^A_n(\omega' - \omega,  \vec k)  + \GG^A_n(\omega' + \omega,  \vec k) \right]  A_m (\omega',  \vec k) .
  \label{alpha}
\end{align}
Further, the term $ \mathfrak G^{nm}_{ij} (\vec k)$ represents the multiorbital quantum-geometric tensor \cite{Ma2010, Kruchkov2023}, defined as  \begin{align}
\mathfrak G^{nm}_{ij}  (\vec k) \equiv \langle \partial_{k_i} u_{n\vec k} | u_{m\vec k}  \rangle \langle u_{m\vec k} |  \partial_{k_j} u_{n\vec k}   \rangle .
\label{metric-nm}
\end{align}
Note that  summing over all other bands $\sum_{m\ne n} \mathfrak G^{nm}_{ij}  = \mathfrak G^{(n)}_{ij}$, is setting the quantum-geometric tensor  $\mathfrak G^{(n)}_{ij}$ of the $n$-th  band, defined by formula
\begin{align}
\mathfrak G^{(n)}_{ij} (\vec k) = \langle \partial_{k_i} u_{n \vec k} | \left [1-      | u_{n \vec k} \rangle \langle u_{n \vec k}  |  \right] | \partial_{k_j} u_{n \vec k} \rangle . 
\label{metric}
\end{align}
The real part of quantum metric tensor $\mathfrak G^{(n)}_{ij} (\vec k)= \mathcal G_{ij} (\vec k) - \frac{i}{2} \eps_{ij} \mathcal F_{xy}$ is \textit{Fubini-Study metric} $\mathcal G_{ij}$ \cite{Provost1980, Fubini1904, Study1905} is responsible for geometry of the manifold,  while the imaginary part is related to Berry curvature $\mathcal F_{xy}$,  responsible for topology.    In ideal flat bands, the quantum geometric tensor's imaginary and real parts are intrinsically connected,  \cite{Kruchkov2022, Wang2021}, see also \cite{Haldane2011}, \footnote{
Equation \eqref{trace-condition} should be understood as a \textit{lower bound} on quantum metric that is fully realized in the case of perfectly flat bands. A deviation from the perfect flatness will result in an augmentation of quantum metrics in left-hand side of equation \eqref{trace-condition}. Nonetheless, in practical materials with nearly-flat bands, such as twisted bilayer graphene, $\Trace \mathcal G_{ij}$ tracks $\mathcal F_{xy}$ quite closely, especially in regions where the band dispersion is the flattest. This is illustrated in Fig. 5 of Ref. \cite{Guan2022}.},  
\begin{align}
\Trace \mathcal G_{ij} (\vec k) = |\mathcal F_{xy} (\vec k)|. 
\label{trace-condition}
\end{align}
We further use this formula with convention for the positively-defined Berry curvature.

Equation \eqref{noise-topo} offers a universal expression for quantum noise in dispersionless quantum states with nontrivial Wannier orbitals. The size and overlaps of electronic orbitals is set by the trace of the quantum metric $\sum_{\vec k} \Trace \mathcal G_{ij} (\vec k)$ \cite{Marzari1997}. As a consequence, in topological insulators the Wannier orbitals cannot be exponentially localized in 2D \cite{Brouder2007, Thouless1984}, hence their quantum geometric tensor shall significantly contribute to the quantum noise via Eqs.\eqref{noise-topo}--\eqref{metric-nm}. Furthermore, the quantum noise \eqref{noise-topo} serves as a probe for the bandgap $\Delta$. To illustrate this link, we examine a minimal model involving two flat Chern bands.

\textit{Two-band dispersionless Chern insulator.}--- We now specifically focus on a two-band model characterized by nearly dispersionless flat topological bands. For instance, such system can be engineered starting from the Haldane model \cite{Haldane1988} by introducing long-range hopping elements that further flatten the bands. By fine-tuning the hopping parameters within the extended hopping range $\Lambda$, the Chern bands of Haldane model can be rendered flat with an exceptional degree of precision \cite{Neupert2011,Kruchkov2022}. Each band in this system is characterized by  the (first) Chern number $C$, a topological invariant which can take any integer values by virtue of expression
\begin{align}
C = \frac{1}{2 \pi} \int_{\rm BZ} d^2 \vec k \, \mathcal F_{xy} (\vec k). 
\label{Chern}
\end{align}
Specifically, in the context of the Haldane model, $C$ acquires values of $\pm 1$. Moreover, our framework allows usage of other flat topological bands with arbitrary high Chern numbers, and Bloch topologies characterized by other topological invariants, such as Euler numbers.  These straightforward models bear relevance to real-world materials, including twisted bilayer graphene and twisted transition metal dichalcogenides, known to feature nearly-flat topological bands \cite{TKV,Khalaf2019,Devakul2021}.

In the case of two flat topological bands, separated by a gap  $\Delta$, the expression for quantum noise \eqref{noise-topo} simplifies, 
\begin{align}
\mathcal S^{\rm flat}_{ij} (\omega)  =  e^2 \Delta^2     \sum_{\vec k}  \lambda_{\vec k} (\omega, \varepsilon_F)   \mathfrak G_{ij} (\vec k) .
\label{noise-2band}
\end{align}
In this context, $\mathfrak G_{ij} (\vec k)$ corresponds to the quantum-geometric tensor \eqref{metric} of the filled band (band index is omitted), and the Fermi level is positioned in the gap.

The key takeaway of formula \eqref{noise-2band} is that the current noise in this case  probes the topological band gap $\Delta$, irrespective of the structure of quasiparticle propagators.  Such probing, however,  relies on complicated frequency dependence in $\lambda_{\vec k} (\omega, \varepsilon_F) $, and requires in-depth knowledge of  interactions in the system and knowledge of exact behavior of the quantum metric  within the Brillouin 
zone $\mathfrak G_{ij} (\vec k)$.
To avoid these complications, we implement the integrated across frequencies current noise, the quantity which serves as an unequivocal probe for the topological band gap.

\textit{Integrated current noise}.---Although the noise function $\lambda_{\vec k} (\omega; \varepsilon_F)$, defined above Eq.\eqref{alpha}, demonstrates a non-universal frequency dependence, its integrated structure can yield a concise analytical result.  Such approach is motivated by experimental methods involving interpretations of quantum noise, where the noise characteristics are often averaged. We proceed with considering noise function $\lambda_{\vec k} (\omega)$ for the case with the Fermi level in the gap. 
 We assume that the Dyson equation for the interacting system
\begin{align}
G_n(\vec k,  \omega) = \frac{1}{\omega - \varepsilon_{n  \vec k} - \Sigma_n (\vec k, \omega)}	,
\end{align}
can be self-consistently resolved  in terms of quasiparticles 
\begin{align}
G_n(\vec k,  \omega) \simeq \frac{Z_{n \vec k}}{\omega - \epsilon_{n\vec k} + i \gamma_{n \vec k} }	 + \text{Regular part},
\end{align}
where the renormalized band dispersion is given by equation  
$\epsilon_{n\vec k} = \eps_{n\vec k} + \Re \, \Sigma_n ( \vec k, \epsilon_{n\vec k}) $. The well-resolved quasiparticles are characterized by 
 the quasiparticle weight $Z_{n \vec k} \sim 1$,  with 
\begin{align}
Z_{n \vec k} = \left[1 - \frac{\partial \Sigma_n (\omega, \vec k)}{\partial \omega} \right]^{-1}	_{\omega = \epsilon_{n \vec k}}, 
\label{residue}
\end{align}
and the quasiparticle (inverse) lifetime is accounted by
\begin{align}
\gamma_{n \vec k} = \Im \Sigma(\epsilon_{n\vec k}, \vec k)	. 
\end{align}
While manifestations of interactions can be rich and diverse in their nature, we presume that interactions in our case preserve the \textit{order of quasiparticle poles}, and keep the bands flat (i.e., $\epsilon_{n \vec k}$$\approx$const), leading to an interacting band gap of $ \Delta_{nm} (\vec k) = \epsilon_{m} (\vec k) -\epsilon_{n} (\vec k) $. Our derivation below shows that interaction-induced renormalization of the pre-existing band gap does not affect the integrated noise.  
Moreover, our derivation does not require $\gamma_{n \vec k}$ to be small, in contrast to Ref.\cite{Neupert2013}; it can be of any reasonable strength and possess $\vec k$-dependence. 
Thereafter the noise function from bands $n$ and $m$ is given by 
\begin{widetext}
\begin{align}
\lambda^{nm}_{\vec k} (\omega, \varepsilon_F) = i  \int \limits_{-\infty}^{\varepsilon_F}   \frac{d \omega'}{\pi}
	 \frac{ Z_{n \vec k} Z_{m \vec k} \gamma_{n \vec k}}
{
(\omega'  - \epsilon_{n \vec k} + i \gamma_{n \vec k})
(\omega'  - \epsilon_{n \vec k} - i \gamma_{n \vec k})
(\omega' - \omega - \epsilon_{m \vec k} + i \gamma_{m \vec k})
}
\nonumber
\\
+
 \frac{ Z_{n \vec k} Z_{m \vec k}  \gamma_{m \vec k}}
{
(\omega'  - \epsilon_{m\vec k} + i \gamma_{m\vec k})
(\omega'  - \epsilon_{m \vec k} - i \gamma_{m\vec k})
(\omega' + \omega - \epsilon_{n \vec k} - i \gamma_{n\vec k})
}  + (\omega  \leftrightarrow - \omega).
\label{lambda}
\end{align}
\end{widetext}
In the expression above we keep the band indices $n,m$ for straightforward generalization to the multiband case. 

The frequency dependence can be estimated through the residue theorem.  For the two-band model with the Fermi level in the gap, we obtain 
\begin{align}
	\lambda^{12}_{\vec k} (\omega) = 2 \pi i  \sum_{\omega'_*} \text{Res} \left[ \alpha^{nm}_{\vec k} (\omega, \omega')  \right] + \text{Regular}. 
\end{align}
where $\alpha^{nm}_{\vec k} (\omega, \omega') $ is the integrand of Eq.\eqref{lambda}, see also definition \eqref{alpha}. 
The relevant poles are given by $\omega'_* = \{\epsilon_{n \vec k} + i \gamma_{n \vec k}, \epsilon_{n \vec k} + \omega + i \gamma_{n \vec k},  \epsilon_{n \vec k} - \omega + i \gamma_{n \vec k}\}$. 
The direct evaluation of residues yields 
\begin{align}
	\lambda^{12}_{\vec k} (\omega)  =   \frac{ 2 i [\Delta_{12} + i \gamma_{12} ]  Z_{1 \vec k} Z_{2 \vec k}  }{\omega^2 - [ \Delta_{12}  + i \gamma_{12}]^2} + \text{Regular}, 
\end{align}
where $\gamma_{nm} = \gamma_{m \vec k} - \gamma_{n \vec k}$.
Note that the main contribution to $\lambda^{12}_{\vec k} (\omega)$ spikes at frequencies in order of band gaps, $|\omega| \approx \Delta_{mn}$. The contribution of the regular term can be neglected when the lower band is fully filled, and upper band is empty. 

In the two-band model, the integrated current noise is linked to the noise function $\lambda_{\vec k} (\omega)$ via Eq. \eqref{noise-2band}.
  Subsequently, the integrated noise for this model becomes
\begin{align}
\int \limits_{-\infty}^{+\infty}  d \omega \, \mathcal S^{\rm flat}_{ij} (\omega)  = e^2 \Delta^2  \sum_{\vec k}  \chi_{\vec k}   \mathfrak G_{ij} (\vec k) ,
\label{noise-2band-b}
\end{align}
where
\begin{align}
\chi_{\vec k} = \int \limits_{-\infty}^{+\infty}  d \omega	\, \lambda_{\vec k} (\omega)  =  2 \pi  \, Z_{1 \vec k} Z_{2 \vec k} .
\label{noise-int}
\end{align}
For systems amenable to a quasiparticle description, the self-energy $\Sigma (\omega, \vec k)$ is a smooth function, and thus the on-shell derivative $\partial \Sigma (\omega, \vec k)/\partial \omega$ remains small. Consequently, the quasiparticle weight, defined in Eq. \eqref{residue}, is close to unity, $Z_{\vec k} \simeq 1$. Therefore, for systems with well-defined quasiparticle, we have
\begin{align}
\chi_{\vec k}   \simeq  2 \pi .
\label{chi}
\end{align}
For a non-interacting system in its clean limit, $ \chi_{\vec k} \equiv 2 \pi $ is a precise statement.

Using formulas \eqref{noise-2band-b} and \eqref{chi}, we can express the integrated quantum noise in ideal flat bands through their topological invariants. In the context of the two-band topological insulator discussed above, we invoke the trace condition linking the Fubini-Study metric with the Berry curvature for ideal flat bands \eqref{trace-condition}. In the leading approximation, the integrated current noise is
\begin{align}
 \int \limits_{-\infty}^{+\infty}  d \omega \, \left[  \mathcal S^{\rm flat}_{xx} (\omega) + \mathcal S^{\rm flat}_{yy} (\omega)  \right]  \simeq  
C  e^2 \Delta^2 .
 \label{band gap 2}
\end{align}
This uses the definition of Chern number $C$ provided in Eq. \eqref{Chern}. 

Equation \eqref{band gap 2} is the main result of our work: It unequivocally links the integrated current noise to the topological band gap $\Delta$. Interaction effects that preserve quasiparticle poles manifest themselves through reduction of quasiparticle weight $Z_{\vec k} \leq 1$. These effects are counterbalanced by the augmentation of quantum metrics deviating from the ideal \eqref{trace-condition} when the condition for perfect band flatness is relaxed. Nevertheless, such corrections remain secondary to the RHS of formula \eqref{band gap 2}.   Hence, equation \eqref{band gap 2} stands as a  robust observable for flat band systems, applicable when the Fermi level is in the gap and the quasiparticle description is meaningful.

\textit{Discussion.}--- 
Our study revisits the current noise as a useful signal in dispersionless quantum states with nontrivial Wannier orbitals. The main findings of our work are threefold:
(i) A comprehensive formula for quantum noise at finite temperature for a system with nontrivial Wannier orbitals and moderate interactions, as presented in Eq. \eqref{current-noise-gen}.
(ii) A low temperature formula for quantum noise in a system with nontrivial Wannier orbitals and at least one dispersionless quantum state, as given by Eq. \eqref{noise-topo}.
(iii) A direct illustration, using a two-orbital system, that the integrated current noise probes the topological band gap, as captured in Eq. \eqref{band gap 2}.

We propose a new method for topological band gap probe:
(i) Carry out a low-temperature measurement of current noise in two channels: $ \mathcal S^{\rm flat}_{xx} (t)$ and $ \mathcal S^{\rm flat}_{yy} (t)$.
(ii) Perform Fourier analysis on the measured data to obtain $ \mathcal S^{\rm flat}_{xx} (\omega)$ and $ \mathcal S^{\rm flat}_{yy} (\omega)$.
(iii) Determine the integrated current noise $\mathfrak S \equiv \int d \omega  \left[ \mathcal S^{\rm flat}_{xx} + \mathcal S^{\rm flat}_{yy} \right]$.
(iv) Evaluate the topological band gap using   formula 
$
\Delta  \approx \frac{1}{e}  \sqrt{\frac{\mathfrak S }{C} } .
$
Estimates suggest accuracy of such method around 10\%-15\%. Strong deviations from these values presume the quasiparticle notion is broken \footnote{Shot noise measurements have been recently implemented to investigate the breakdown of quasiparticle concept  \cite{Chen2022}.}.

Twisted transition metal dichalcogenides (TMDs) are good examples of materials with flat topological bands.  Recent experimental work has detected signatures of (anomalous) fractional Chern insulators in the flat bands of twisted TMDs \cite{ Cai2023, Park2023, Zeng2023,Xu2023}. In this context, a natural extension of such experiments would be application of quantum noise for direct probing of fractional electric charge in these materials \cite{Kane1994,FQHcharge1,FQHcharge2,FQHEcharge3}. This makes twisted TMDs  a promising platform for noise probing of topological band gaps in dispersionless quantum states.

\

\textit{Acknowledgements.} The authors thank Titus Neupert and Daniil Evtushinsky for useful discussion.  A.K. is supported by the Branco Weiss Society in Science, ETH Zurich, through the grant on flat bands, strong interactions, and the SYK physics. 
S.R.~is supported by the National Science Foundation under 
Award No.\ DMR-2001181, and by a Simons Investigator Grant from
the Simons Foundation (Award No.~566116).
This work is supported by
the Gordon and Betty Moore Foundation through Grant
GBMF8685 toward the Princeton theory program.

\bibliography{Refs}

\end{document}